\documentclass[12pt,a4paper]{article}
\usepackage{subfigure}   
\usepackage{graphicx}

\begin{document}

\begin{center}
{\LARGE Planar Thirring
Model in the U(2$N$)-symmetric limit}

\vskip 1cm
{\bf Simon Hands}

\vskip 0.5cm
{\em Department of Physics, Faculty of Science \& Engineering,\\Swansea University,\\
Singleton Park, Swansea SA2 8PP,United Kingdom. }

\end{center}

\begin{abstract}
I review the Thirring model in 2+1$d$ dimensions, focussing in particular on
possible strongly-interacting UV-stable fixed points of the
renormalisation group, corresponding to a continuous phase transition where a
U($2N$) global symmetry spontaneously breaks to U($N)\otimes$U($N$).
Since there is no small parameter in play, a systematic non-perturbative
approach such as numerical simulation of lattice field theory is
mandated. I compare and contrast various formulations, paying particular attention
to models formulated with either staggered or domain wall lattice
fermions. Domain wall fermions, which faithfully capture U($2N$) symmetry in the
limit of wall separation $L_s\to\infty$, predict a critical flavor number
$1<N_c<2$.
\end{abstract}




\section{Introduction}\label{sec:intro}
Most of our understanding of quantum field theory derives from theories
in which interactions are relatively weak, and it makes sense to envisage
elementary processes in terms of well-localised particles. A powerful
diagrammatic calculus has been developed to treat this situation both
quantitively and conceptually -- the shorthand  for this being  ``perturbation
theory''. Even in cases such as the strong interaction where perturbation 
theory fails to capture the esseential physics, our calculational
techniques, often the numerical simulation of euclidean lattice field theory,
rely on there being a well-understood path to a perturbative continuum limit;
for QCD this is guaranteed by asymptotic freedom.

In general, neither weak interactions nor well-localised particle degrees
of freedom are necessary ingredients for a continuum quantum field theory.
Our modern perspective is that QFTs exist as fixed points of the
renormalisation group, where the ratio of the  physical scale $\mu$ to some UV
regulator scale $\Lambda$ can be made arbitrarily small by a suitable tuning
of parameters $\beta\to\beta^*$, such that
$(\mu/\Lambda)\propto(\beta-\beta^*)^\nu$, where $\nu>0$ is one of a set of
interrelated {\em
critical exponents\/} characterising the continuum theory. 
In this way predictions are essentially independent of the
regularisation details. Even within this framework it can be
challenging to calculate accurately, especially in the absence of a small
dimensionless expansion parameter. Lattice field theory simulation
relies on no small parameters; however, there may still be barriers to complete control 
to be overcome, particularly for theories involving fermions. The
lattice provides a natural regularisation for quantities expressible in
terms of differential forms, but theories containing a single relativistic
fermion species do not fall in this class~\cite{Rabin:1981qj}.

\section{The Thirring Model}
The Thirring model in 2+1$d$ has Lagrangian density  
\begin{equation}
{\cal
L}=\bar\psi_i(\partial\!\!\!/\,+m)\psi_i+{{g^2}\over{2N}}(\bar\psi_i\gamma_\mu\psi_i)^2.
\label{eq:Thirring}
\end{equation}
Here $\psi$ is a spinor in a reducible representation of the Clifford
algebra, meaning that it is acted on by $4\times4$ Dirac matrices $\gamma_\mu$
satisfying $\{\gamma_\mu,\gamma_\nu\}=2\delta_{\mu\nu}$ in euclidean metric. The index $i=1,\dots
N$ runs over $N$ distinct species of relativistic fermion. The interaction between
conserved current densities $i\bar\psi\gamma_\mu\psi$ implies, as
in electrodynamics, that opposite charges attract, like charges repel.
Some example motivations for a model of this genre 
are the description of nodal fermions in $d$-wave
superconductors~\cite{Tesanovic:2002zz,Herbut:2002yq}, spin-liquid phases of
Heisenberg antiferromagnets~\cite{Wen:2002zz, Rantner:2002zz}, certain 
correlated Chern insulators amenable to quantum simulation with ultra-cold
atoms~\cite{Ziegler:2020zkq},
and low-energy
electronic excitations in graphene~\cite{CastroNeto:2009zz,Hands:2008id}.

Our approach to understanding the model (\ref{eq:Thirring}) at strong coupling
is rooted in its symmetries. In 2+1$d$ the reducible representation of the Dirac
algebra contains two elements $\gamma_3$ and
$\gamma_5\equiv\gamma_0\gamma_1\gamma_2\gamma_3$ which anticommute
with the kinetic term in (\ref{eq:Thirring}). For mass $m\to0$, the following 
spinor rotations generate a U(2$N$) symmetry 
\footnote{To demonstrate this
using infinitesimal rotations it is convenient to define $\tilde\psi=\bar\psi
i\gamma_3\gamma_5$}
with $\alpha_i$ a
Hermitian generator of U($N$):
\begin{eqnarray}
\psi\mapsto e^{i\alpha_1}\psi;\;\bar\psi\mapsto\bar\psi e^{-i\alpha_1}&:&\;\;\;
\psi\mapsto e^{\alpha_{35}\gamma_3\gamma_5}\psi;\;\bar\psi\mapsto\bar\psi
e^{-\alpha_{35}\gamma_3\gamma_5};\label{eq:unbroken}\\
\psi\mapsto e^{i\alpha_3\gamma_3}\psi;\;\bar\psi\mapsto\bar\psi
e^{i\alpha_3\gamma_3}&:&\;\;\;
\psi\mapsto e^{i\alpha_{5}\gamma_5}\psi;\;\bar\psi\mapsto\bar\psi
e^{i\alpha_5\gamma_5}.\label{eq:broken}
\end{eqnarray}
Once $m\not=0$, only (\ref{eq:unbroken}) remain valid; in a gapped theory the
symmetry is therefore broken to U($N)\otimes$U($N$).
A second important symmetry frequently motivated by
condensed-matter applications is symmetry  under inversion of spatial axes --
in a particle physics context this is equivalent to {\em parity}, arising from
inversion of an odd number of euclidean axes; our convention is
$x_\mu\mapsto-x_\mu$, $\mu=0,1,2$. For reducible spinors there are two
inequivalent parity flips:
\begin{eqnarray}
P_3:\;\;\psi(x)\mapsto\gamma_3\psi(-x)&;&\;\;\;\bar\psi(x)\mapsto\bar\psi(-x)\gamma_3\\
P_5:\;\;\psi(x)\mapsto\gamma_5\psi(-x)&;&\;\;\;\bar\psi(x)\mapsto\bar\psi(-x)\gamma_5.
\end{eqnarray}

We thus identify the full global symmetry of (\ref{eq:Thirring}) with $m=0$ as
U($2N)\otimes$Z$_2\otimes$Z$_2$. Three mass terms break
U($2N)\to$U$(N)\otimes$U($N$):
\begin{equation}
++\;\;m_h\bar\psi\psi;\;\;\;
+-\;\;im_3\bar\psi\gamma_3\psi;\;\;\;
-+\;\;im_5\bar\psi\gamma_5\psi,
\label{eq:masses}
\end{equation}
where the $\pm$ indicate parity  under $P_{3,5}$ flips. 
In graphene, the $++$ mass corresponds to a charge density wave in which electrons 
preferentially sit on one of two sub-lattices on the
honeycomb~\cite{Semenoff:1984dq},  
whereas a linear combination of $+-$ and $-+$ yields a bond density wave in which 
electrons are distributed on both sublattices in a Kekul\'e texture~\cite{Hou:2006qc}.
Spontaneous breaking of the continuous 
symmetry is accompanied by $2N^2$ Goldstones, $N^2$ with 
$J^P=0^{-}$ and $N^2$ $J^P=0^{+}$, with $P$ the unbroken parity.
Finally, note that there is another possible mass term:
\begin{equation}
--\;\;m_H\bar\psi\gamma_3\gamma_5\psi,
\end{equation}
not related to (\ref{eq:masses}) by any U($2N$) rotation. This is
the non-time-reversal invariant {\em Haldane mass\/} realised in 
graphene-like systems by alternately circulating currents in adjacent half-unit
cells~\cite{Haldane:1988zza}; we will not consider it further. 

The Thirring model is often analysed by introducing a vectorlike auxliary
boson field $A_\mu$, and (\ref{eq:Thirring}) replaced by the equivalent form  
\begin{equation}
{\cal L}^\prime=\bar\psi_i(\partial\!\!\!/\,
+i{g\over\surd N}A\!\!\!\!\;\!/\,+m)\psi_i+\textstyle{1\over2}A_\mu A_\mu.
\label{eq:Thirringaux}
\end{equation}
In an expansion in powers of $N^{-1}$, the leading order quantum correction
arises in the auxiliary two-point  function $D_{\mu\nu}$, due to a diagram analogous
to vacuum polarisation in electrodynamics. The result is UV-finite  if the
regularisation respects current conservation~\cite{Gomes:1990ed,Hands:1994kb}:
\begin{equation}
D_{\mu\nu}(k)={{{\cal P}_{\mu\nu}(k)}\over{1-\Pi(k^2)}}+{{k_\mu k_\nu}\over k^2},
\label{eq:DA}
\end{equation}
where ${\cal P}_{\mu\nu}(k)=\delta_{\mu\nu}-k_\mu k_\nu/k^2$ is the
transverse projector, and in $d$ spacetime dimensions
\begin{equation}
\Pi(k^2)=-g^2{{4\Gamma(2-{d\over2})}\over{3(4\pi)^{d\over2}m^{4-d}}}k^2F(2;2-{\textstyle{d\over2}};
{\textstyle{5\over2}};
-{k^2\over4m^2}).
\end{equation}
Asymptotically,
\begin{equation}
\lim_{k^2\to\infty}\Pi(k^2)=-g^2{{(k^2)^{{d\over2}-1}}\over A_d}
\label{eq:UV}
\end{equation}
with $A_d$ a numerical constant taking the value 8 for $d=3$. Now, if we
restrict attention to $A\bar\psi\psi$ vertices involving a conserved fermion current, as
in a gauge theory, then the longitudinal component of (\ref{eq:DA}) is physically
irrelevant, and hence the UV asymptotics of diagrams contributing to higher order
corrections, following (\ref{eq:UV}), is controlled by $D_{\mu\nu}\propto{\cal
P}_{\mu\nu}/g^2k^{d-2}$. The outcome is that the Thirring model is power-counting
renormalisable in a continuous range of dimensions $d\in(2,4)$, with only
logarithmic
divergences which are absorbed in fermion wavefunction and mass
renormalisations. The coupling $g^2$ requires no further renormalisation,
the model being governed by the dimensionless
combination  $m^{d-2}g^2$. A key factor in assessing the model's dynamics
is the ratio $M_V/m$ where the mass of the vector bound state  $M_V$ is given by
the physical pole of (\ref{eq:DA})~\cite{Hands:1994kb}. For weak couplings the vector is a
weakly-bound $\psi\bar\psi$ state and for $d=3$ at leading order in $N^{-1}$,
\begin{equation}
{M_V\over m}=2\left(1-2e^{-{2\pi\over mg^2}}\right),
\end{equation}
to be contrasted with the strong coupling result 
\begin{equation}
{M_V\over m}=\sqrt{{6\pi}\over{mg^2}}.
\label{eq:MVlargeN}
\end{equation}

This unexpected renormalisability may not be the end of the story -- the $1/N$
expansion has nothing to say about possible ground states in which U(2$N$)
is spontaneously broken by the generation of a bilinear condensate such as
$\langle\bar\psi\psi\rangle\not=0$ and the consequent dynamical generation
of a massgap $\Sigma$, essentially because all Feynman diagrams
contributing to the signal vanish due to Furry's theorem, just as in QED.
\footnote{However, the $N^{-1}$ expansion in $d=2+\varepsilon$ 
has been used to predict a UV fixed point~\cite{Hikami:1976at}.}
However, non-perturbative approaches suggest bilinear condensation may
occur for sufficiently large $g^2$ and/or small $N$. For instance
self-consistent solution of the Schwinger-Dyson (SD) equation,
using the bifurcation method and approximating the full vector propagator by
(\ref{eq:DA}), finds $\Sigma\not=0$ in the strong-coupling limit
$g^2\to\infty$ for $N<N_c=128/3\pi^2\simeq4.32$~\cite{Itoh:1994cr}. The critical line 
extends into the plane with $g_c^2(N<N_c)$ a monotonically decreasing
function~\cite{Sugiura:1996xk}.  Alternative 
approximations to the exact SD equations find
$N_c=32/\pi^2\simeq3.24$~\cite{Gomes:1990ed}, $N_c=2$~\cite{Kondo:1995np}, or even
$N_c=\infty$~\cite{Hong:1993qk}.

In the context of graphene, the existence of a symmetry-breaking phase
transition separating
a conducting semi-metal from a Mott insulator has technological
significance; if undoped graphene were an insulator 
the fabrication of fast graphene-based switches with a stable ``off''
state would be possible, with the potential for a new generation of fast
processors. For monolayer graphene the physical value $N=2$
4-spinors, corresponding to 2 atoms/unit cell on the honeycomb lattice $\times$
2 distinct ``Dirac points'' (where the gap closes linearly) in the first Brillouin Zone
$\times$ 2 electron spin states.
Unfortunately simulations of graphene using a realistic Hamiltonian
appear to preclude this possibility~\cite{Ulybyshev:2013swa}. However, the
question also has intrinsic theoretical interest. If the transition is
second-order as suggested by the self-consistent approaches, it suggests
a correlation length $\xi\sim\Lambda/\Sigma$ diverging as $g^2\to g_c^2(N)$
and the possibility  of an interacting theory in the continuum
limit, described by a UV-stable RG fixed point as set out above. 
We could also describe this as a Quantum Critical Point, with a distinct 
QCP for every integer $N<N_c$.
Since there are
no small dimensionless parameters in play, elucidation of the QCP
properties, and even determining 
the value of $N_c$, requires essentially non-perturbative calculational techniques.
In addition to SD, the Functional Renormalisation Group (FRG) has
also been applied to this question~\cite{Gies:2010st}.
In the remainder of this article, however, I will focus on lattice field theory.

\section{The Staggered Thirring Model}
In any question of dynamical fermion mass generation, the natural starting point
on the lattice is the staggered formulation, involving single-component Grassmann
fields $\chi,\bar\chi$ living on the sites $x$ of a cubic lattice. These are technically
straightforward to implement and have a U($N)\otimes$U($N$) global symmetry
protecting massless fermions from acquiring a gap in perturbation theory. 
The version of the Thirring model which has been most studied has
action~\cite{DelDebbio:1997dv}
\begin{eqnarray}
S_{stagg}&=&{1\over2}\sum_{x\mu i}
\bar\chi_x^i\eta_{\mu x}(1+iA_{\mu x})\chi_{x+\hat\mu}^i
-\bar\chi_x^i\eta_{\mu x}(1-iA_{\mu x-\hat\mu})\chi_{x-\hat\mu}^i\nonumber\\
&+&m\sum_x\bar\chi_x^i\chi_x^i+{N\over{4g^2}}\sum_{x\mu}A^2_{\mu x}.
\label{eq:Sstagg}
\end{eqnarray}
This is recognisably of the same form as (\ref{eq:Thirringaux}) with Dirac
matrices replaced by Kawamoto-Smit phases $\eta_{\mu
x}\equiv(-1)^{x_0+\cdots+x_{\mu-1}}$. In the absence of interactions staggered
fermions recover the continuum action of $N_f$ reducible flavors in
the long-wavelength limit, with $N_f=2N$~\cite{Burden:1986by}. The vector auxiliary field $A_\mu$
is defined on the links joining the sites, just like a lattice gauge field,
with the important distinction that the resulting link field
is unbounded in magnitude and accordingly not unitary. Lattice formulations with
compact auxiliary fields have also been
explored~\cite{Kim:1996xza,Alexandru:2016ejd,Narayanan:2021rcu}. The formulation
(\ref{eq:Sstagg}) has the virtue that integration over $A_\mu$ recovers a
fermion interaction term containing no higher than four-point couplings. 

\begin{figure}
\centerline{\includegraphics[width=5.8cm]{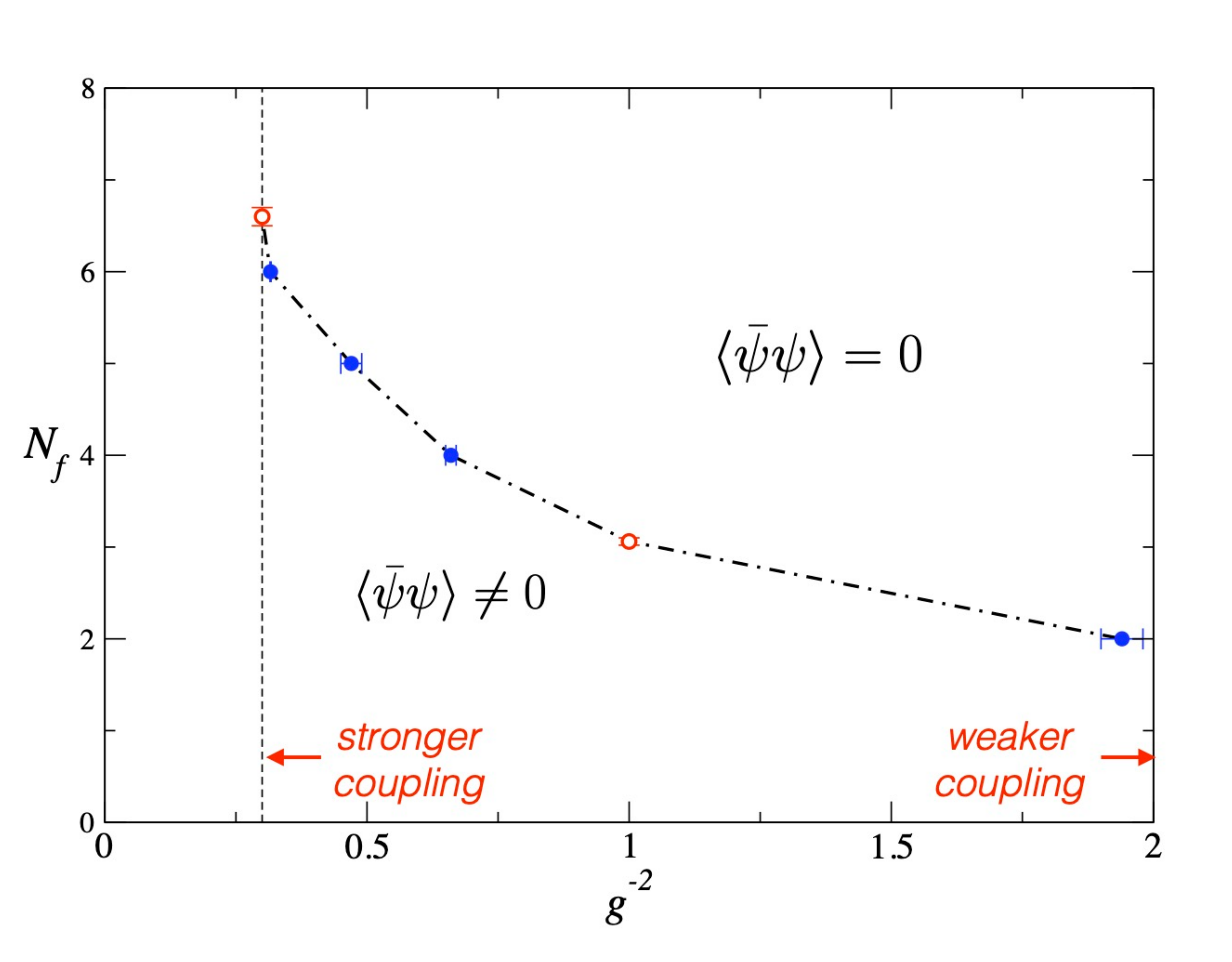}}
\caption{Phase diagram of the staggered lattice Thirring model.} \label{fig:staggeredpd}
\end{figure}
The phase diagram obtained using the staggered lattice action (\ref{eq:Sstagg})
showing the boundary between semimetal and insulator phases, 
is shown in Fig.~\ref{fig:staggeredpd}. Non-integer values of $N$ were  simulated
using a hybrid R algorithm which employs a version of rooting. The figure also
contains a point obtained in the strong-coupling limit~\cite{Christofi:2007ye}
corresponding to a critical flavor number 
\begin{equation}
N_{fc}=6.6(1)
\label{eq:Nfcstagg}
\end{equation}
which is at least of the same order as the SD estimates. 
In contrast to SD predictions however, the critical properties of the QCPs
are very sensitive to $N$; in particular 
the 
exponent $\delta$ characterising the order parameter response to a
small bare mass at criticality varies from $\delta=2.75(9)$ at
$N_f=2$~\cite{DelDebbio:1997dv}, to $\delta=6.90(3)$ at
$N_f=N_{fc}$~\cite{Christofi:2007ye}.
The
vertical dashed line marks the estimated location of the strong coupling
limit, which does not coincide with $g^{-2}=0$. Empirically it corresponds to
the location of a maximum in the order parameter data
$\langle\bar\psi\psi(g^2)\rangle$, which is unexpectedly
non-monotonic. A possible origin~\cite{DelDebbio:1997dv} is that
the lattice action
(\ref{eq:Sstagg}) does not have an exactly conserved fermion current, as a consequence of
the non-compactness of the link variable. In the $N^{-1}$ expansion this results
in a divergent contribution $\sim g^2a^{-1}\delta_{\mu\nu}$ to the vacuum
polarisation, which must be removed by an additive renormalisation of $g^{-2}$.
The unphysical region to the left with $g_R^{-2}<0$ has $D_{\mu\nu}$ negative,
corresponding to a violation of positivity. 

Overall, the strong $N$-dependence of the staggered Thirring model is striking,
and the qualitative resemblance of Fig.~\ref{fig:staggeredpd} to SD
predictions~\cite{Sugiura:1996xk} is encouraging. However, the case is not yet
closed. The minimal  model (\ref{eq:Sstagg}) with $N=1$ (corresponding to
$N_f=2$ in Fig.~\ref{fig:staggeredpd}) has been studied using a fermion bag
algorithm~\cite{Chandrasekharan:2011mn}, 
which unlike the workhorse HMC algorithm used elsewhere permits Monte Carlo sampling 
in the massless limit $m=0$, and hence a much cleaner estimate of critical
properties. The exponents obtained are compatible with the older HMC
values~\cite{DelDebbio:1997dv}; however they are also indistinguishable from
those of the minimal 2+1$d$ Gross-Neveu model with U(1) global
symmetry~\cite{Chandrasekharan:2013aya}. When the lattice action is expressed
purely in terms of fermion  fields $\chi,\bar\chi$ this is not surprising:
interactions occur among staggered fields distributed around the vertices of an
elementary cubic cell, and the two models differ only  by a (presumably
irrelevant) body-diagonal term. Only once  auxiliary bosons are introduced, as
in both $N^{-1}$ expansion and HMC algorithm, do the models look different.
Since the GN auxiliary is scalar/pseudoscalar, symmetry breaking corresponds to 
auxiliary condensation, is captured by the $N^{-1}$
expansion~\cite{Hands:1992be}, and occurs for all $N$. While
Fig.~\ref{fig:staggeredpd} suggests the staggered Thirring and GN models
for arbitrary $N$ are distinct, it is {\em a priori} hard to understand why the
corresponding continuum models should coincide for $N$ minimal.

The resolution is that the symmetry-breaking pattern of the staggered
model (\ref{eq:Sstagg}) is U($N)\otimes$U($N)\to$U($N$), whereas that of
the continuum model (\ref{eq:Thirringaux}) is U($2N)\to$U($N)\otimes$U($N$).
Only in  the weak coupling limit is the continuum form recovered using
$N_f=2N$~\cite{Burden:1986by}; at a generic QCP we cannot expect the ``taste
symmetry restoration'' anticipated in the continuum limit of lattice QCD, amd must conclude that the
QCPs of the staggered lattice Thirring model lie in a different universality
class to the continuum model (\ref{eq:Thirring}).
The consequences are in principle profound, as the following non-rigorous
argument demonstrates. Observe that according to the large-$N$ prediction
(\ref{eq:MVlargeN}), the mass of the vector boson vanishes in the strong coupling
limit. At this point the Thirring model becomes a theory of conserved currents
interacting via exchange of a massless vector, ie. QED$_3$, so it
is plausible that the critical $N_c$ required for symmetry breaking coincides
with $N_c$ defining the IR conformal fixed point of QED$_3$~\cite{Appelquist:1986fd}.
Next, there is an old conjecture~\cite{Appelquist:1999hr} that symmetry breaking in any theory is
constrained by the relation
\begin{equation}
f_{IR}\leq f_{UV},
\label{eq:finequality}
\end{equation}
where $f$ is proportional to the negative of the thermodynamic free energy
density, and essentially counts the number of light degrees of freedom.
For an asymptotically-free theory like QED$_3$, the count in the UV is the number of spinor degrees 
of freedom multiplied by a factor ${3\over4}$ associated with Fermi-Dirac statistics in 2+1$d$.
Assuming a gap-generating symmetry breaking, the IR count is the number of Goldstones, 
$N^2$ for U($N)\otimes$U($N)\to$U($N$), and $2N^2$ for U($2N)\to$U($N)\otimes$U($N$). 
In either limit the count is supplemented by 1 for the photon, which is massless in each phase.
Application of (\ref{eq:finequality}) then predicts
\begin{eqnarray}
N^2\leq{3\over4}\times2^3N\;&\Rightarrow&\;N_c\leq6\;\Rightarrow\:N_{fc}=12\;\;\;\mbox{staggered};\label{eq:Nstagg}\\
2N^2\leq{3\over4}\times4N\;&\Rightarrow&\;N_c\leq{3\over2}\;\;\;\;\;\;\;\;\;\;\;\;\;\;\;\;\;\;\;\;\;\;
\mbox{continuum}.\label{eq:Ncont}
\end{eqnarray}
The big disparity between (\ref{eq:Nstagg}) and (\ref{eq:Ncont}) hints that capture of the correct symmetry is
key to correctly modelling the QCP and predicting $N_c$.

\section{The DWF Thirring Model: Formulation}
In recent years, in an attempt to overcome the shortcomings of the staggered model, two altenative lattice fermion 
formulations have been investigated. One uses the SLAC
derivative~\cite{Drell:1976mj}, in which global symmetries are manifest while the fermion kinetic bilinear is
non-local to suppress the influence of species doublers at the Brillouin Zone
edge. The usual arguments that the SLAC derivative leads to non-covariant divergent
terms~\cite{Karsten:1979wh} do not apply in this case because there is no
gauge-invariance requirement forcing a delocalised fermion-boson interaction; rather the  
auxiliary fields $A_\mu$ are defined on sites~\cite{Wellegehausen:2017goy}. 
The other approach is motivated by the resemblance of the fermion-boson
interaction in (\ref{eq:Thirringaux}) to that of a gauge theory, and uses the
domain wall fermion (DWF) originally developed to accurately capture flavor symmetries
in the chiral limit of lattice QCD.

DWF are formulated on a 2+1+1$d$ lattice in which open boundaries separated by a
distance $L_s$ are imposed in the fictitious third direction $x_3$. 
If we write the Lagrangian density
\begin{equation}
{\cal L}_{\rm DWF}=\bar\Psi(x,s)D^{\rm DWF}_{x,y;s,s^\prime}\Psi(y,s^\prime),
\label{eq:DWF}
\end{equation}
with $s$ the $x_3$-coordinate,
then as $L_s\to\infty$, near zero-modes of $D^{\rm DWF}$ are localised on
``domain walls'' at $x_3=0,L_s$ as approximate $\pm$ eigenmodes of
$\gamma_3$. Physical fields in the 2+1$d$ target space are then defined by
\begin{equation}
\psi(x)=P_-\Psi(x,1)+P_+\Psi(x,L_s);\;\;\;
\bar\psi(x)=\bar\Psi(x,L_s)P_-+\bar\Psi(x,1)P_+,
\label{eq:physical}
\end{equation}
with projectors $P_\pm={1\over2}(1\pm\gamma_3)$. 
As originally shown by
Kaplan~\cite{Kaplan:1992bt}, if $D^{\rm DWF}$ has the
same form as the Wilson lattice derivative, then species doublers are not
present in the low-energy spectrum. What about the U($2N$) symmetry?
While the physical degrees of freedom have
readily-identifiable components which are eigenstates of $\gamma_3$, the full
symmetry generated by (\ref{eq:unbroken},\ref{eq:broken}) requires 
that ${\cal L}_{\rm DWF}$ also has well-controlled behaviour under field
rotations generated by $\gamma_5$.
The resolution, mirroring the steps originally developed for theories in 3+1$d$,
is that $D^{\rm DWF}$ is a finite-$L_s$ regularisation of
a fermion operator~\cite{Ginsparg:1981bj} $D$ satisfying the {\em
Ginsparg-Wilson\/} relations
\begin{equation}
\{\gamma_3,D\}=2D\gamma_3D;\;\;\;
\{\gamma_5,D\}=2D\gamma_5D;\;\;\;
[\gamma_3\gamma_5,D]=0.
\label{eq:GW}
\end{equation}
Formally the RHS of the GW relations (\ref{eq:GW}) is $O(aD)$, so U($2N$)
is recovered in the long-wavelength limit provided $D$ is sufficiently
local~\cite{Hands:2015qha}. An overlap
operator~\cite{Neuberger:1997fp,Neuberger:1998wv,Hands:2015dyp} 
$D^{\rm ov}$ satisfying
(\ref{eq:GW}) but not manifestly local in 2+1$d$ can be constructed. 

DWF mass terms only involve fields defined on the walls. The equivalents of
(\ref{eq:masses}) read:
\begin{eqnarray}
m_h[\bar\Psi(x,L_s)P_-\Psi(x,1)&+&\bar\Psi(x,1)P_+\Psi(x,L_s)]\nonumber\\
im_3[\bar\Psi(x,L_s)\gamma_3P_-\Psi(x,1)&+&\bar\Psi(x,1)\gamma_3P_+\Psi(x,L_s)]\label{eq:DWFmasses}\\
im_5[\bar\Psi(x,1)\gamma_5P_-\Psi(x,1)&+&\bar\Psi(x,L_s)\gamma_5P_+\Psi(x,L_s)].\nonumber
\end{eqnarray}
Note that while $m_{h,3}$ couple fields on opposite walls as in 3+1$d$, $m_5$
couples fields on the {\em same\/} wall. Nonetheless, pilot studies with quenched
QED$_3$ show that bilinear condensates of the same form as (\ref{eq:DWFmasses})
coincide in the limit $L_s\to\infty$ as required by U($2N$) symmetry, and
moreover the 3 and 5 condensates are numerically
indistinguishable~\cite{Hands:2015qha}. Significantly, finite-$L_s$ corrections
are considerably smaller in the 3,5 channels than for $h$; accordingly in
subsequent work we focus on $i\langle\bar\psi\gamma_3\psi\rangle$ rather than
$\langle\bar\psi\psi\rangle$. When simulating full fermion dynamics with DWF, 
the impact of modes propagating in the bulk $0<x_3<L_s$ must be decoupled using
suitably chosen bosonic Pauli-Villars fields~\cite{Furman:1994ky}. For a 2+1$d$
theory the key relation is~\cite{Hands:2015dyp}
\begin{equation}
{{\mbox{det}D^{\rm DWF}(m_i)}\over{\mbox{det}D^{\rm
DWF}(m_h=1)}}=\mbox{det}D_{L_s}(m_i)\;\;\;\mbox{with}\;\;\;
\lim_{L_s\to\infty}D_{L_s}(m_i)=D^{\rm ov}(m_i).
\label{eq:GWlimit}
\end{equation}

Even after restricting our consideration to formulations with a non-compact auxiliary
link field,
there are still different plausible choices for the DWF-auxiliary interaction
term~\cite{Hands:2016foa}. The so-called {\em Surface\/}
formulation confines the interaction to $\Psi,\bar\Psi$ defined on the walls,
ie.
\begin{equation}
S_{\rm surf}={i\over2}\sum_{x\mu} A_{\mu x}[\bar\Psi_{x1}\gamma_\mu
P_-\Psi_{x+\hat\mu1}+\bar\Psi_{xL_s}\gamma_\mu P_+\Psi_{x+\hat\mu L_s}]-{\rm
h.c.}
\end{equation}
This is by analogy with the treatment of the scalar auxiliary in the Gross-Neveu
model formulated with DWF, simulation of which captures a continuous
symmetry-breaking phase transition perfectly
adequately~\cite{Vranas:1999nx,Hands:2016foa}. A technical advantage is that the
Pauli-Villars determinant in the denominator of (\ref{eq:GWlimit}) does not depend on $A_\mu$, so can be ignored in the
simulation. By contrast the {\em Bulk} formulation treats $1+iA_\mu$ as
a gauge connection, albeit non-unitary. Accordingly the auxiliary
interacts with all fermion fields including those in the bulk:
\begin{equation}
S_{\rm bulk}={i\over2}\sum_{x\mu s}A_{\mu
x}[\bar\Psi_{xs}(-1+\gamma_\mu)\Psi_{x+\hat\mu s}]+A_{\mu x-\hat\mu}
[\bar\Psi_{xs}(1+\gamma_\mu)\Psi_{x-\hat\mu s}].
\end{equation}
The bulk formulation is much more costly to simulate, in part due to to the
non-unitary nature of the link fields, and in part because in the minimal
$N=1$ model there is a slight obstruction to proving positivity of the fermion
determinant,  so the results to be reviewed below require the RHMC algorithm
simulating functional weight $\mbox{det}(M^\dagger
M)^{1\over2}$~\cite{Hands:2018vrd}. 
Schematically writing $D^{\rm DWF}=D_{\rm W}+D_3$, where $D_{\rm W}$ resembles an
orthodox 2+1$d$ Wilson fermion derivative, then 
it's important to note that the property 
$\gamma_3D_{\rm W}\gamma_3=\gamma_5D_{\rm W}\gamma_5=D_{\rm W}^\dagger$ needed
in the demonstration~\cite{Hands:2015dyp}  of (\ref{eq:GWlimit}) does not
require unitary link fields.

\section{The DWF Thirring Model: Numerical Results}
\begin{figure}
\centerline{\includegraphics[width=6.8cm]{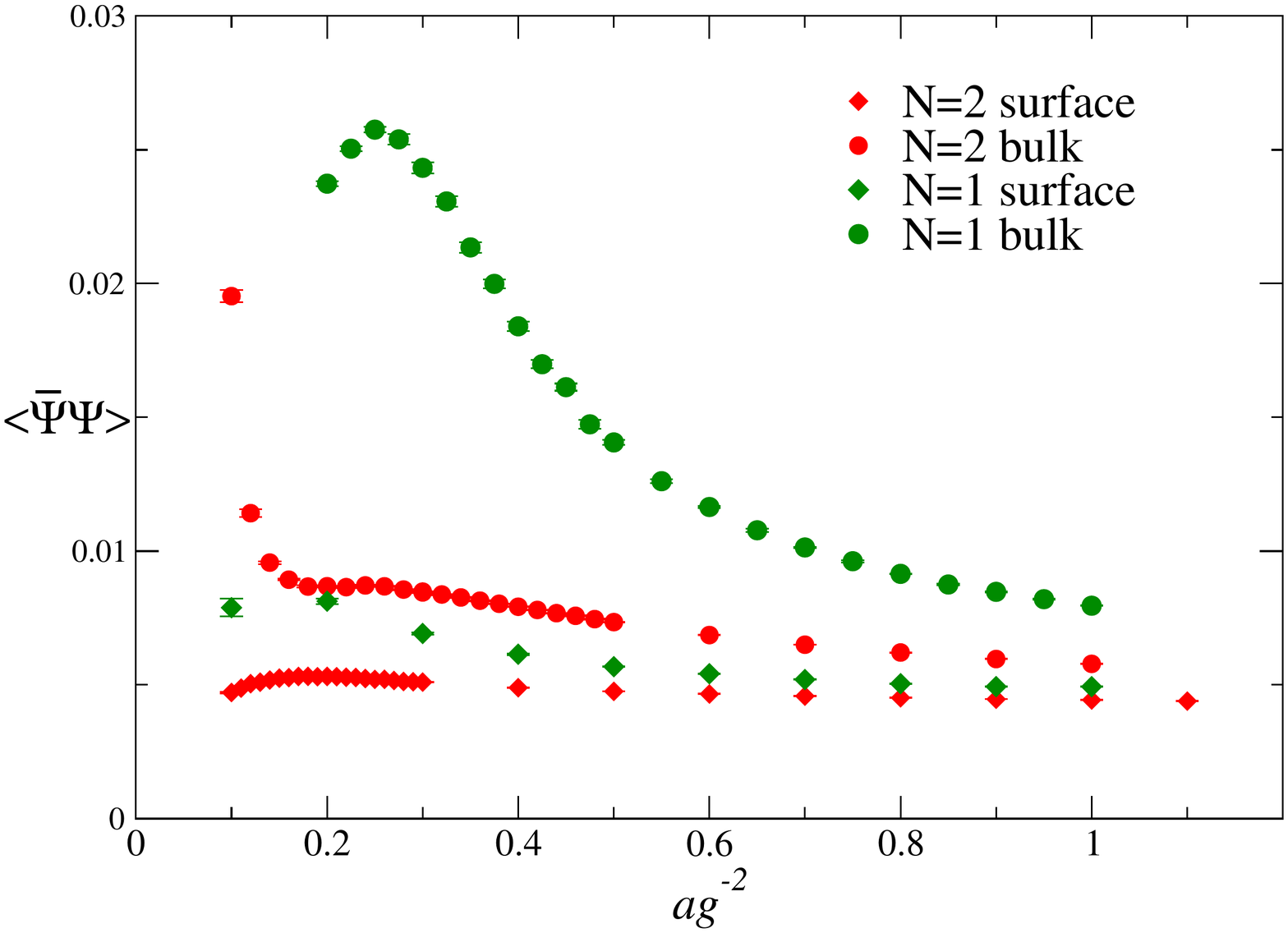}}
\caption{Bilinear condensate as a function of inverse coupling.}
\label{fig:cond_N}
\end{figure}
We now review results of simulations with the Thirring model defined with DWF
fermions. Fig.~\ref{fig:cond_N} shows the bilinear condensate
$\langle\bar\psi\psi\rangle$ (actually
$i\langle\bar\psi\gamma_3\psi\rangle$ as commented above) with $ma=0.01$ 
on a $12^3\times L_s$ as a function of inverse
coupling for models with $N=2$ ($L_s=16$)~\cite{Hands:2016foa}, and $N=1$
($L_s=8$)~\cite{Hands:2018vrd}, using both surface and bulk formulations. 
The bulk signal is much larger than that for surface, and for $N=1$ over
$N=2$; indeed for all cases examined with $N=2$,
$\langle\bar\psi\psi\rangle$ varies linearly with fermion mass $m$ implying
U($2N$) symmetry remains unbroken~\cite{Hands:2018vrd}. The non-monotonic
variation of the data  suggests 
strong-coupling artifacts just as found for the staggered model.

\begin{figure}[ht]
\centerline{
  \subfigure[$\delta_h(\beta,m)$ on $12^3\times L_s$]
     {\includegraphics[width=2.0in]{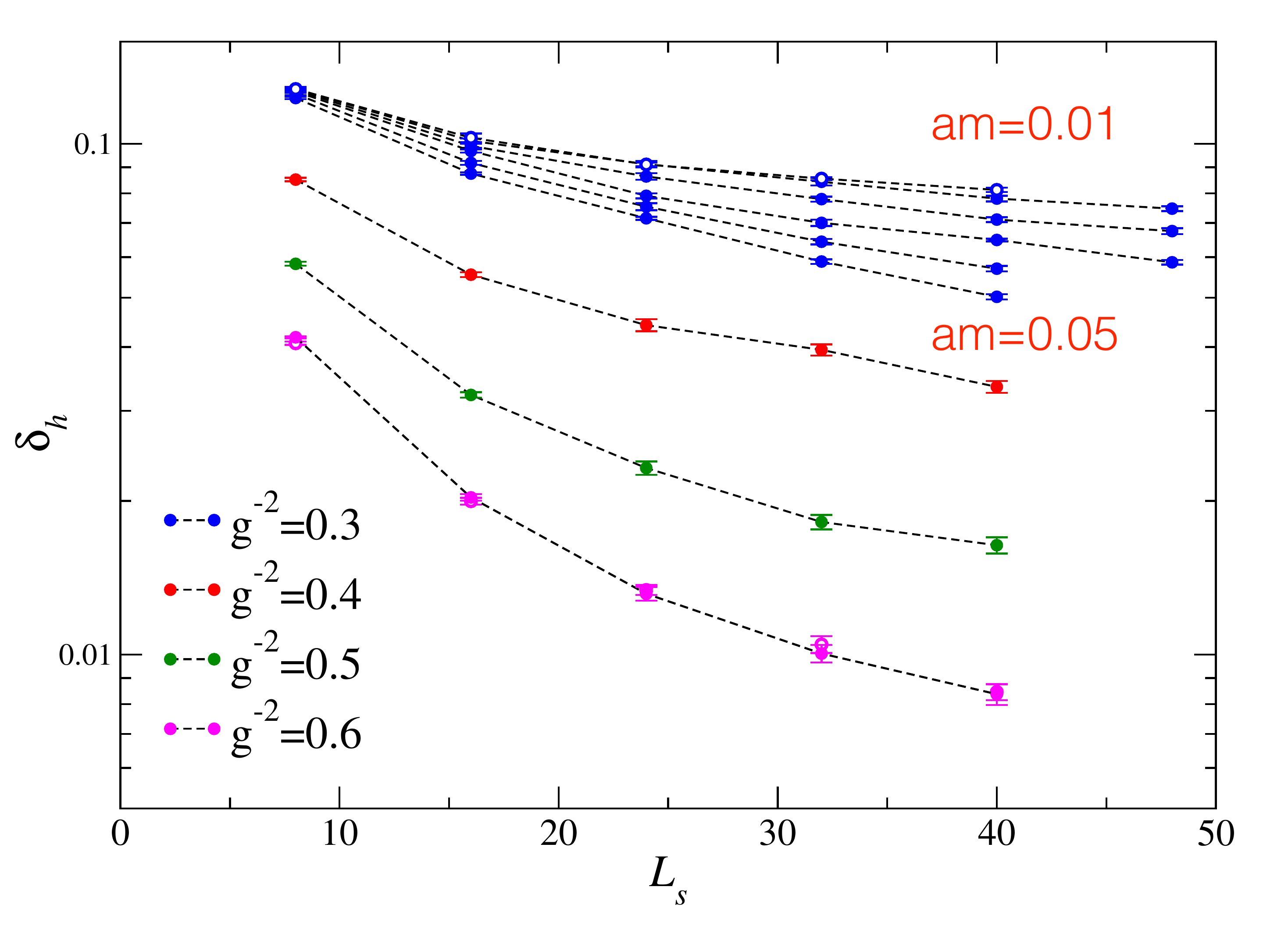}\label{fig:deltah_Ls}}
  \hspace*{4pt}
  \subfigure[$\delta_h(\beta,m)$ on $16^3\times48$]
     {\includegraphics[width=2.1in]{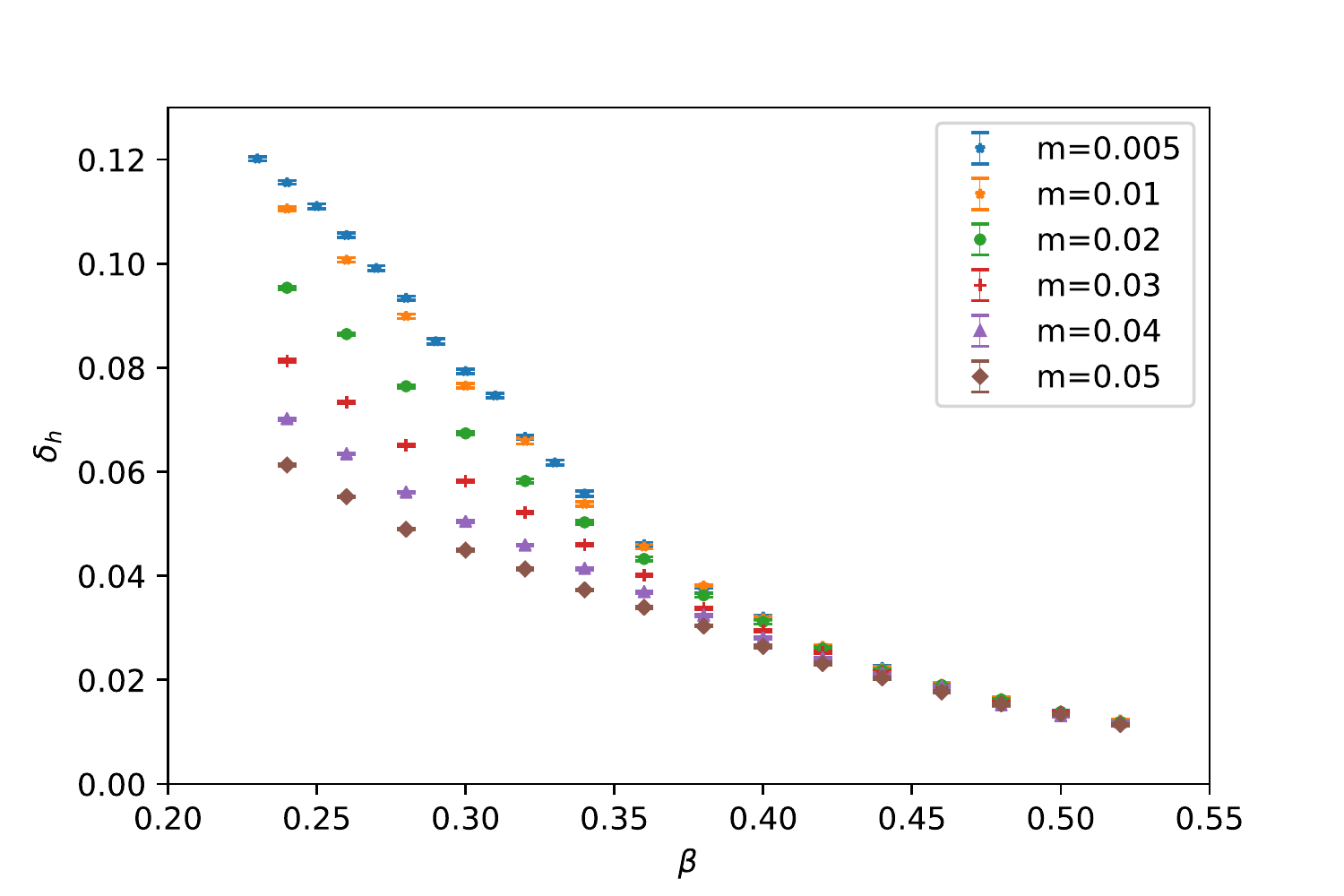}\label{fig:deltah_beta}}
}
\caption{} 
\label{fig:Lsextr} 
\end{figure}

The extrapolation $L_s\to\infty$ is crucial in identifying any possible symmetry
breaking. Empirically data is well-fitted by an exponential  {\em Ansatz}:
\begin{equation}
\langle\bar\psi\psi\rangle_{L_s\to\infty}-\langle\bar\psi\psi\rangle_{L_s}=
C(m,g^2)e^{-\Delta(m,g^2)L_s},
\label{eq:Lsfit}
\end{equation}
which we have tested up to $L_s=48$, where the decay constant $\Delta\sim
O(10^{-2})$ at the strongest couplings~\cite{Hands:2020itv}.
It is also important to test the recovery of U($2N$) in the same limit,
using the dominant residual
\begin{equation}
\delta_h(L_s)=\mbox{Im}\langle\bar\Psi(x,1)i\gamma_3\Psi(x,L_s)\rangle=
-\mbox{Im}\langle\bar\Psi(x,L_s)i\gamma_3\Psi(x,1)\rangle,
\end{equation}
which empirically matches
$\langle\bar\psi(1-i\gamma_3)\psi\rangle$ closely~\cite{Hands:2015qha}.
This is shown as a function of $L_s$ in Fig.~\ref{fig:deltah_Ls} and of inverse
coupling $\beta=ag^{-2}$ at fixed $L_s=48$ in Fig.~\ref{fig:deltah_beta}.
It appears that $\delta_h$ dwindles away only very slowly as $L_s\to\infty$; moreover
at stronger couplings the residual also grows as $m\to0$. 
The DWF approach to the GW limit defined by (\ref{eq:GW}) will be
numerically challenging.

As in any numerical approach involving an auxiliary field on a finite system, we are hampered
because the bilinear condensate order parameter vanishes identically for $m\to0$, and need to base
a search for a critical point on data generated with $m\not=0$. The approach is
to fit $\langle\bar\psi\psi(g^2,m)\rangle$ to a renormalisation group-inspired
equation of state~\cite{DelDebbio:1997dv}
\begin{equation}
m=A(\beta-\beta_c)\langle\bar\psi\psi\rangle^{\delta-{1\over\beta_m}}+B\langle\bar\psi\psi\rangle^\delta,
\label{eq:EoS}
\end{equation}
where $\beta_m$ and $\delta$ can be identified with conventional critical
exponents: 
\begin{equation}
\langle\bar\psi\psi\rangle\propto(\beta_c-\beta)^{\beta_m};\;\;\;
\langle\bar\psi\psi(\beta_c)\rangle\propto m^{1\over\delta}.
\end{equation}
\begin{figure}
\centerline{\includegraphics[width=6.8cm]{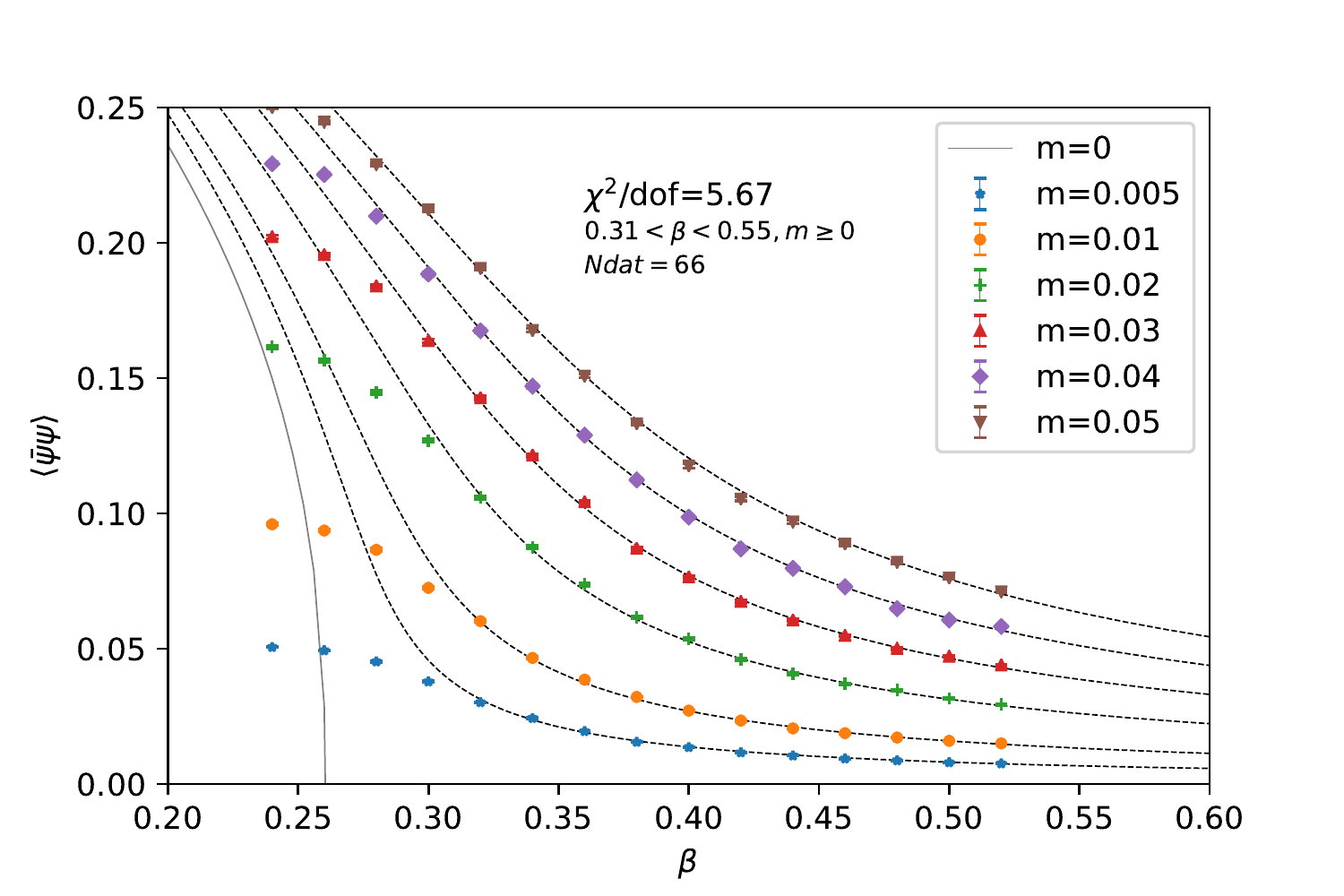}}
\caption{Fits to the Equation of State (\ref{eq:EoS}) on $16^3\times48$.}
\label{fig:eosfit_sc}
\end{figure}
Fig.~\ref{fig:eosfit_sc} shows such a fit from a $16^3\times48$
system. The fitted data has $\beta\in[0.32,0.52]$ and
$ma\in[0.005,0.05]$, and the resulting critical parameters are
$\beta_c=0.2601(4)$, $\beta_m=0.413(15)$ and $\delta=3.44(9)$. These values are
compatible with fits to data extrapolated to $L_s\to\infty$ using
(\ref{eq:Lsfit}), which are theoretically better-founded, but technically more
challenging~\cite{Hands:2020itv}. 
\begin{figure}[ht]
\centerline{
  \subfigure[Critical coupling $\beta_c(L_s)$]
     {\includegraphics[width=2.2in]{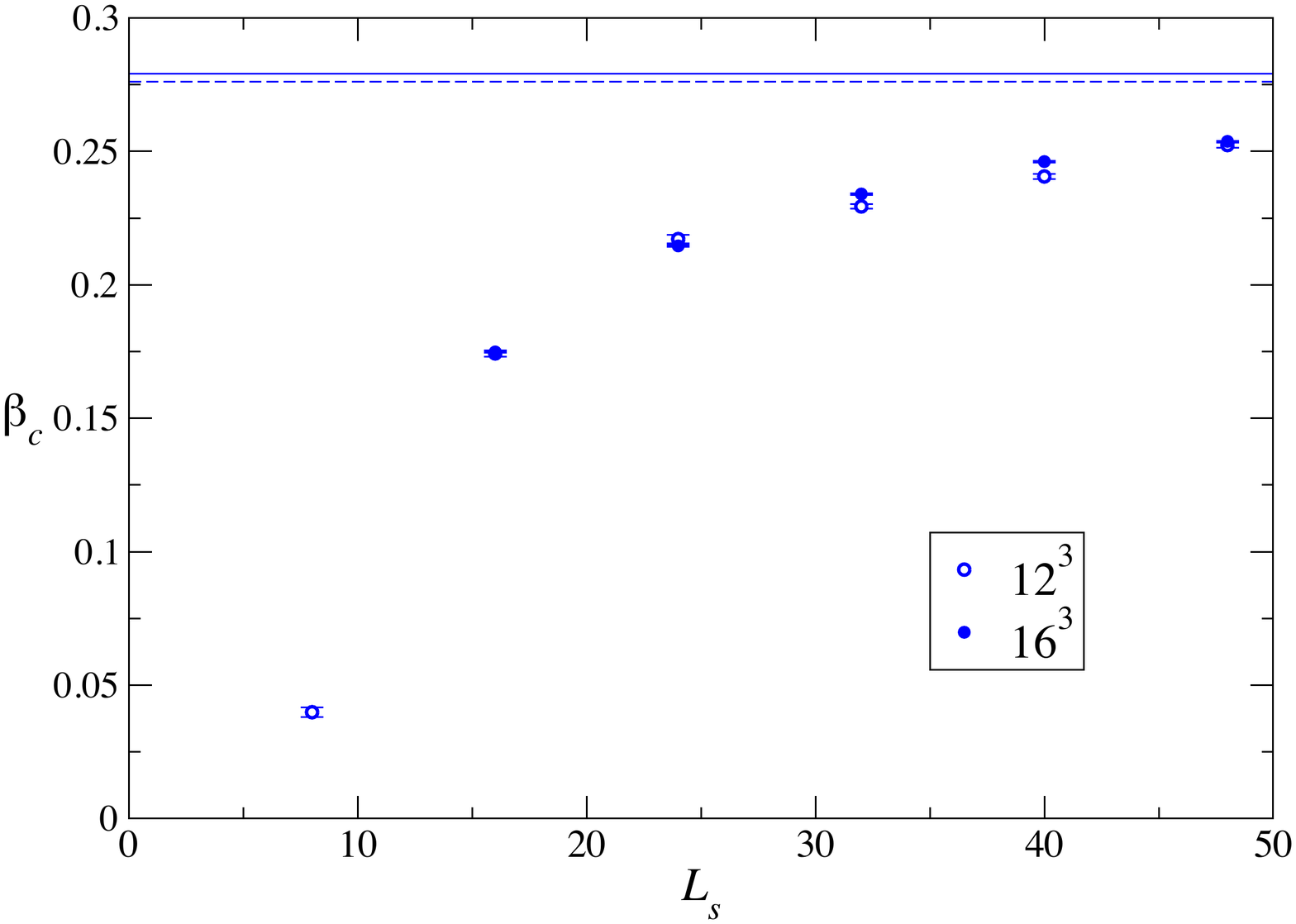}\label{fig:betac_Ls}}
  \hspace*{4pt}
  \subfigure[Exponents $\beta_m(L_s)$ and $\delta(L_s)$]
     {\includegraphics[width=2.2in]{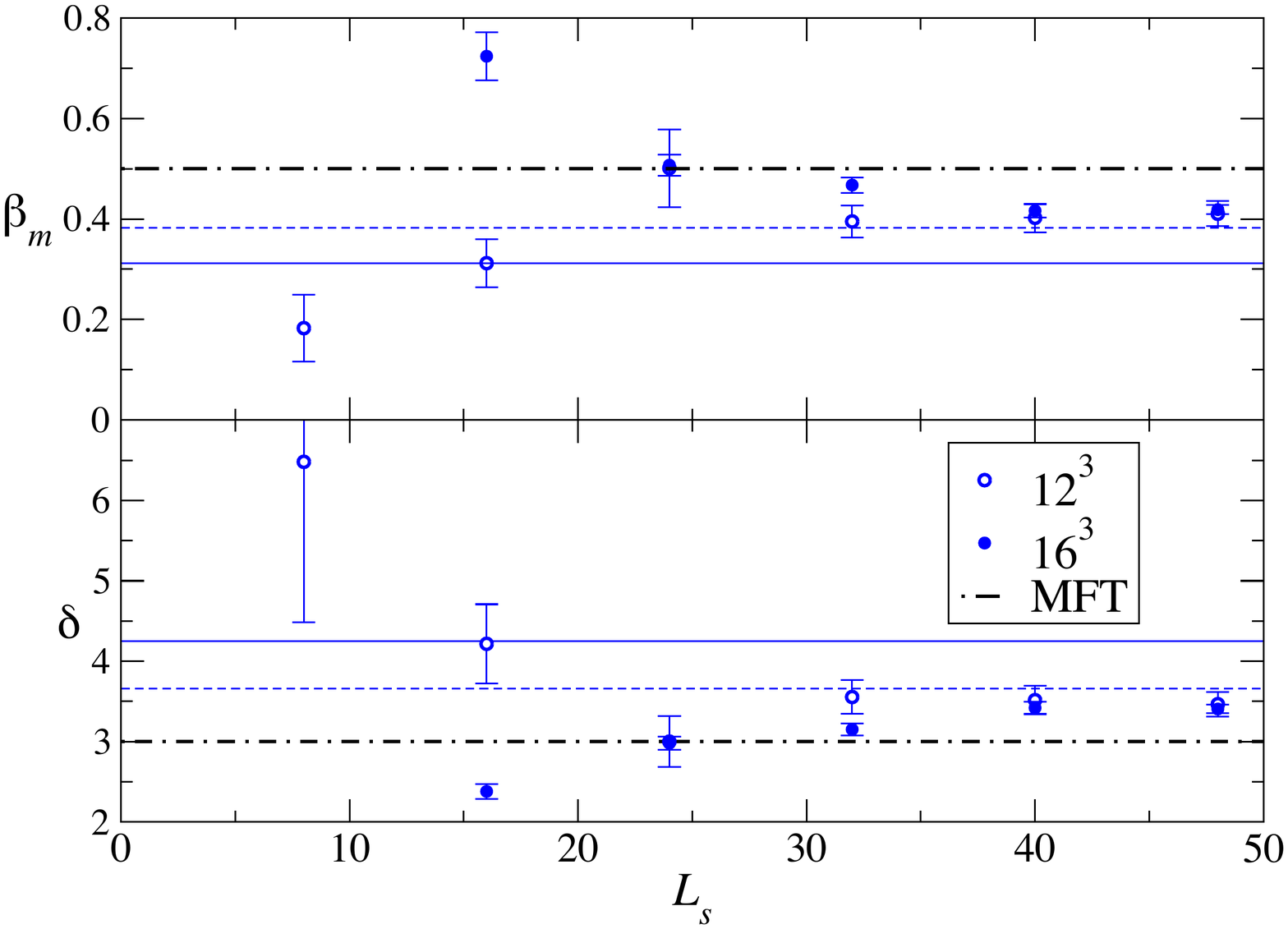}\label{fig:exponents_Ls}}
}
\caption{Critical parameters at finite $L_s$ compared with $L_s\to\infty$.} \label{fig:critfits} 
\end{figure}
A comparison of EoS-fitted critical parameters at fixed $L_s$ with results of
the $L_s$-extrapolated data is shown for $\beta_c$ in Fig.~\ref{fig:betac_Ls} and
for the exponents in Fig.~\ref{fig:exponents_Ls}. It is also apparent that
finite volume corrections assessed via comparison of system sizes $12^3$ and $16^3$
are rather small as $L_s\to\infty$.

In summary, while Fig.~\ref{fig:eosfit_sc} shows data at strong couplings,
including from the hypothesised 
U($2N$)-broken phase, are not well-described by the EoS (\ref{eq:EoS})
(we ascribe the mismatch to large finite-$L_s$ corrections in the broken phase, which 
an ongoing simulation campaign is aiming to 
control), there is by now rather strong evidence that U($2N$) is broken
spontaneously at
strong couplings for $N=1$. The absence of evidence for broken symmetry for
$N=2$~\cite{Hands:2016foa,Hands:2018vrd} leads to the prediction
\begin{equation}
1<N_c<2.
\end{equation}
The contrast with the staggered result (\ref{eq:Nfcstagg}) fully justifies the
investment in DWF. The universality class of the QCP is characterised by
critical exponents with estimated values~\cite{Hands:2020itv} as $L_s\to\infty$
\begin{equation}
\beta_m=0.31(2);\;\;\;\delta=4.3(2).
\end{equation}
These differ significantly from the values $\beta_m=0.57(2)$,
$\delta=2.75(9)$~\cite{DelDebbio:1997dv} or $\beta_m=0.70(1)$,
$\delta=2.64(3)$~\cite{Chandrasekharan:2011mn}
found for the minimal staggered model.
So-called {\em hyperscaling\/} relations can be derived on the assumption that
dynamics near a QCP is characterised by a single diverging length scale
$\xi$. If this is also applicable to fermion models~\cite{Hands:1992be}, then we
can further deduce values for the exponent $\nu$ defined by
$\xi\propto\vert\beta_c-\beta\vert^{-\nu}$ and $\eta$ describing critical
correlations of the order parameter field
$\langle\bar\psi\psi(0)\bar\psi\psi(r)\rangle\propto r^{-(d-2+\eta)}$. The
outcome is
\begin{equation}
\nu={1\over
d}\beta_m(\delta+1)=0.55(3);\;\;\;\eta={{(d+2)-(d-2)\delta}\over{\delta+1}}=0.13(4),
\end{equation}
again in clear distinction to the minimal staggered Thirring
model~\cite{Chandrasekharan:2011mn}. This is evidence that the DWF model with $N=1$ 
supports a novel QCP.

\begin{figure}[ht]
\centerline{
  \subfigure[Localisation of $D^{\rm ov}$ on $16^3$]
     {\includegraphics[width=2.2in]{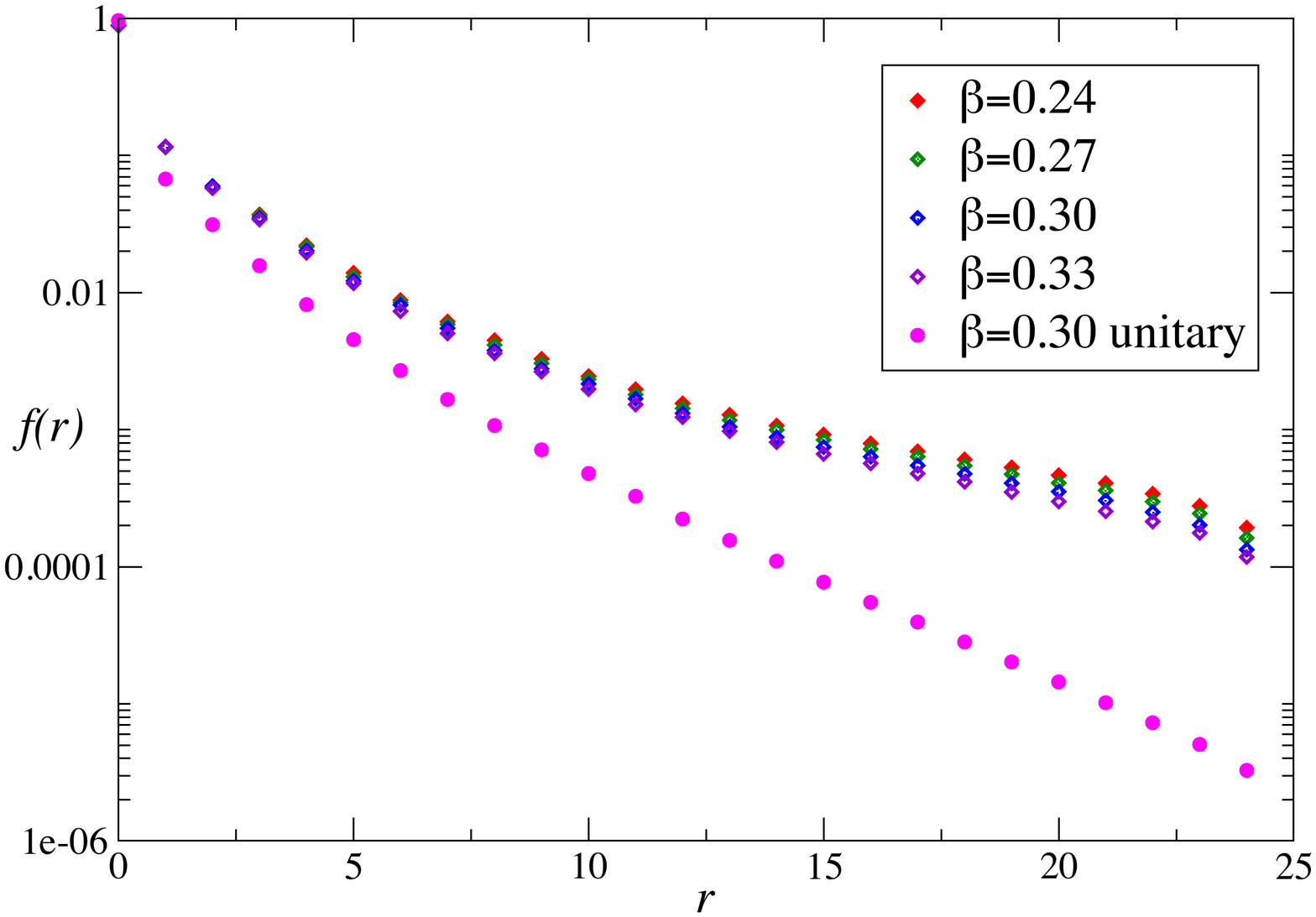}\label{fig:locShamM005g3}}
  \hspace*{4pt}
  \subfigure[Truncation error $\delta_{GW}(L_s)$ for $D^{\rm ov}$]
     {\includegraphics[width=2.2in]{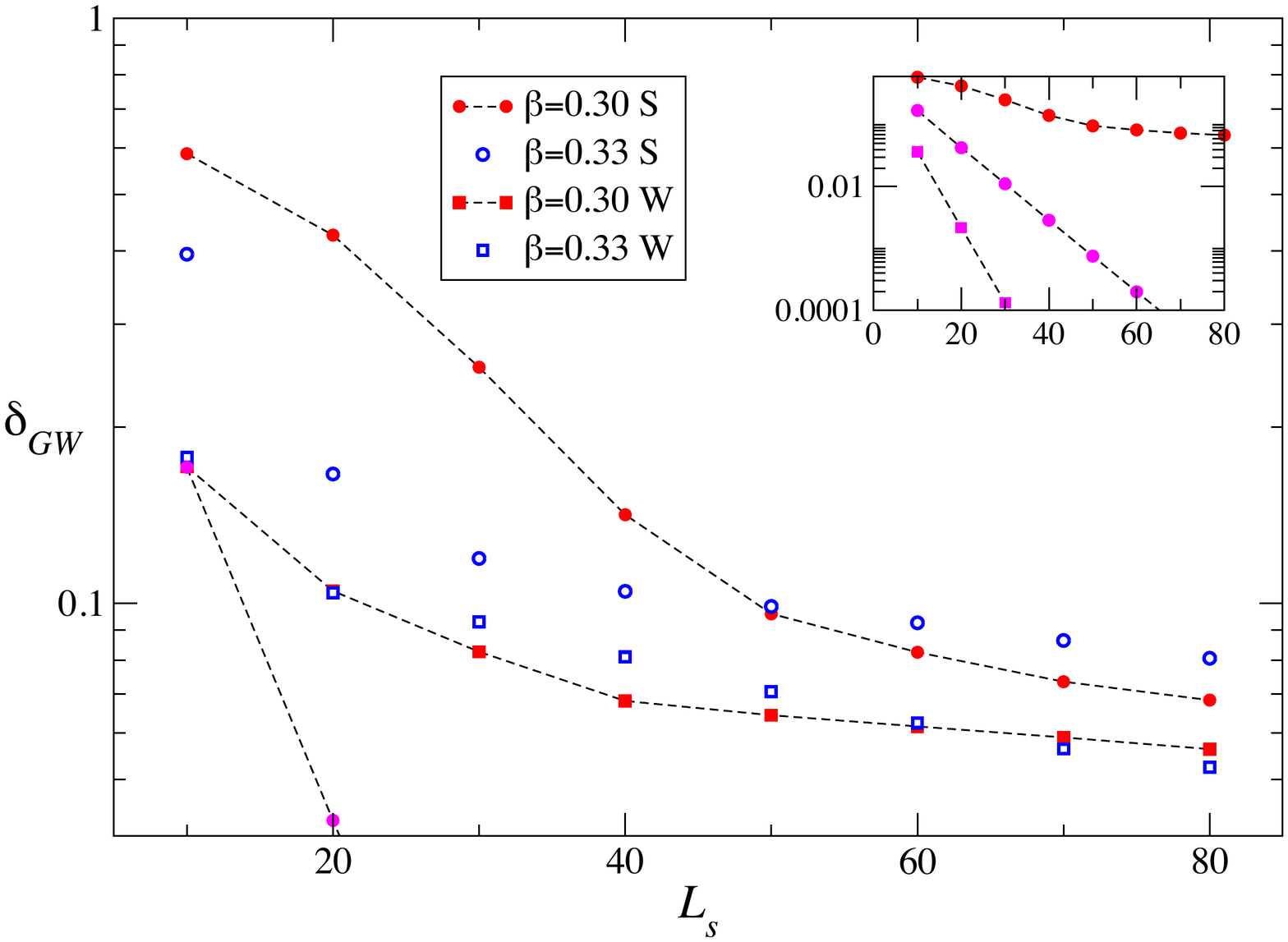}\label{fig:GWerror}}
}
\caption{Tests of locality and GW symmetry (\ref{eq:GW}) for the overlap operator.} 
\label{fig:localisation} 
\end{figure}
In order to ascertain that the GW symmetry (\ref{eq:GW}) actually coincides with
U($2N$) it is necessary to establish the locality of the overlap operator
$D^{\rm ov}$
associated with our DWF implementation. Following a similar analysis for lattice
QCD~\cite{Hernandez:1998et},  Fig.~\ref{fig:locShamM005g3}~\cite{Hands:2020itv} plots 
${\rm max}\{\vert\vert\phi(x)\vert\vert_2:\vert\vert x-y\vert\vert_1=r\}$ for a
vector $\phi$ obtained for various $\beta$ via $\phi=D^{\rm ov}\eta$ with $\eta$ a delta-function
source sited at $y$. The plot confirms exponential localisation of $D^{\rm ov}$
throughout the critical region, as required. Also shown is the same result
obtained using unitary link fields, showing that in this case the localisation
length is significantly smaller. Fig.~\ref{fig:GWerror} from the same work shows
the discrepancy $\delta_{GW}=\vert\vert(\gamma_3D^{\rm ov}+D^{\rm
ov}\gamma_3-2D^{\rm ov}\gamma_3D^{\rm ov})\phi\vert\vert_\infty$ as a function
of $L_s$; like Fig.~\ref{fig:deltah_Ls} it shows the desired symmetry
(\ref{eq:GW}) being
restored, but only rather slowly, as $L_s\to\infty$. Results are shown for both
the Shamir kernel correspondiing to DWF, and the Wilson kernel which performs
slightly better. Again, the inset shows that using unitary link fields improves
matters markedly. 

\section{Discussion}
In summary, the Thirring model defined using the DWF bulk
formulation appears to sustain a QCP where a novel strongly-interacting quantum
field theory may be found in the continuum limit.  The QCP is essentially 
defined by dimensionality, count of light fermion degrees of freedom, and
pattern of global symmetry breaking.
Not even all approaches satisfying these criteria are equivalent: DWF with the
surface formulation of the Thirring interaction yield
$0<N_c<1$~\cite{Hands:2018vrd}, and SLAC fermions predict $N_c=0.80(4)$~\cite{Lenz:2019qwu}. 
In such cases a unitary QFT with integer $N$ cannot
exist.
We speculate that bulk DWF fall in a
different RG basin of attraction due to their promotion of strong dynamics
resembling those of abelian gauge theory: similar reasoning underlies the
conjectured identification with IR QED$_3$ outlined above, and is supported by
the large-$N$ approach. 
It is clear from the results reviewed above that
the 2+1$d$ Thirring model  furnishes a challenging new environment in which to stress-test DWF and assess their
merit as a non-perturbative specification of massless fermions. 

We end with some
open questions. The critical exponents $\beta_m,\delta,\eta,\nu$ studied above
all relate to critical fluctuations of the order parameter field, which is a
fermion bilinear. Understanding strong dynamics in the symmetric phase also
requires the exponent $\eta_\psi$ defined by the critical propagator
$\langle\psi(0)\bar\psi(r)\rangle\propto r^{-(2+\eta_\psi)}$. Unfortunately,
exploratory studies of the DWF fermion propagator have proved very
noisy~\cite{Hands:2016foa}, and appear to require source smearing, in contrast
to the relatively clean results obtained with staggered
fermions~\cite{DelDebbio:1997dv}. In the meson sector there is
evidence for a spectral separation beteween Goldstone and non-Goldstone channels
at strong coupling~\cite{Hands:2018vrd} but lattices with a longer temporal
extent are needed for precision. Finally the ratio
$\langle\psi\psi\rangle/m\chi_\pi$, where $\chi_\pi$ is the integrated Goldstone
correlator (also known as the transverse susceptibility), differs
markedly from unity as the coupling strength
grows~\cite{Hands:2016foa,Hands:2018vrd} contradicting the
axial Ward Identity associated with U($2N$).
The problem may lie in unexpectedly strong parameter renormalisation,  and also 
possibly in the identification (\ref{eq:physical}) of the physical fields; in a
strongly-interacting world it is not clear how firmly tethered
the low-energy degrees of freedom are to the walls.

\begin{figure}
\centerline{\includegraphics[width=6.8cm]{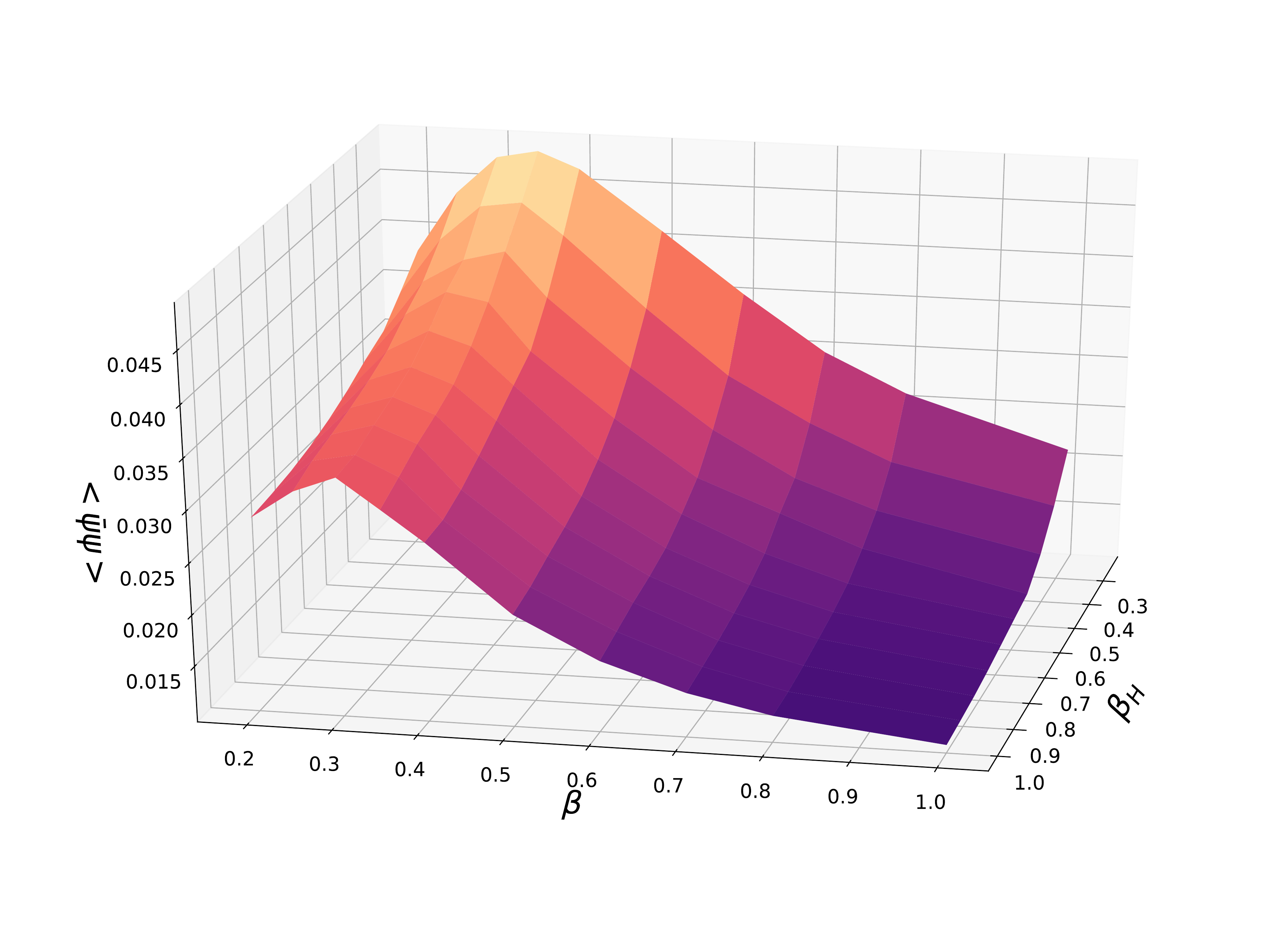}}
\caption{Bilinear condensate as a function of $\beta$, $\beta_H$ on
$12^3\times8$.}
\label{fig:haldane}
\end{figure}
Since the QCP is defined by its symmetry, it also behoves us to consider other
possible interactions consistent with U($2N)\otimes$Z$_2$. It has been
observed~\cite{Gehring:2015vja} that beyond the current-current interaction of
(\ref{eq:Thirring}) there is another contact interaction, the so-called 
Haldane interaction $-g_H^2/2N(\bar\psi\gamma_3\gamma_5\psi)^2$ that shares this
symmetry (the sign of the coupling is determined by the requirement for the
fermion determinant to be positive definite), and accordingly there is no justification to exclude it from the
putative fixed point action; indeed it has been studied using both 
SLAC fermions~\cite{Schmidt:2016rtz} and FRG. Fig.~\ref{fig:haldane} shows results
of pilot DWF simulations with a Haldane interaction localised on the
walls~\cite{BenPage}; 
decreasing $\beta_H\equiv g_H^{-2}$
significantly enhances the
bilinear condensate. There is every prospect that the Haldane interaction is present in
the fixed-point action, as predicted by FRG~\cite{Gehring:2015vja}.

Finally, what is to be made of the staggered lattice model?
In principle it should remain  interesting -- a model exhibiting a sequence of
QCPs defining strongly-interacting QFTs whose universal properties are exquisitely sensitive
to the parameter $N$. The real question is what is $N$ counting? Is it possible
to formulate a continuum quantum field theory of fermions with symmetry breaking
U($N)\otimes$U($N)\to$U($N$)? Can this theory be phrased in terms of local
fields and interactions? 
Perhaps the Thirring model still has some
surprises in store.

\section*{Acknowledgements}
Recent numerical work employed the Cambridge Service for Data Driven Discovery
(CSD3), part of which is operated by the University of Cambridge Research
Computing on behalf of the STFC DiRAC HPC Facility (www.dirac.ac.uk). The
DiRAC component of CSD3 was funded by BEIS capital funding via STFC capital
grants ST/P002307/1 and ST/R002452/1 and STFC operations grant ST/R00689X/1.
DiRAC is part of the National e-Infrastructure. Additional work utilised the
Sunbird facility of Supercomputing Wales. The work was also supported by
STFC grant ST/T000813/1. It is a pleasure to thank my
principal collaborators Luigi del Debbio, Costas Strouthos, Michele Mesiti, Jude
Worthy and Ben Page.

\bibliographystyle{ws-rv-van}


\begin{thebibliography}{10}

\bibitem{Rabin:1981qj}
J.M.~Rabin,
Nucl. Phys. B \textbf{201} (1982), 315-332

\bibitem{Tesanovic:2002zz}
Z.~Te\v sanovi\'c, O.~Vafek and M.~Franz,
Phys. Rev. B \textbf{65} (2002), 180511.

\bibitem{Herbut:2002yq}
I.F.~Herbut,
Phys. Rev. B \textbf{66} (2002), 094504.

\bibitem{Wen:2002zz}
X.G.~Wen,
Phys. Rev. B \textbf{65} (2002), 165113.

\bibitem{Rantner:2002zz}
W.~Rantner and X.G.~Wen,
Phys. Rev. B \textbf{66} (2002), 144501.

\bibitem{Ziegler:2020zkq}
L.~Ziegler, E.~Tirrito, M.~Lewenstein, S.~Hands and A.~Bermudez,
[arXiv:2011.08744 [cond-mat.quant-gas]].

\bibitem{CastroNeto:2009zz}
A.H.~Castro Neto, F.~Guinea, N.M.R.~Peres, K.S.~Novoselov and A.K.~Geim,
Rev. Mod. Phys. \textbf{81} (2009), 109-162.

\bibitem{Hands:2008id}
S.~Hands and C.~Strouthos,
Phys. Rev. B \textbf{78} (2008), 165423.

\bibitem{Semenoff:1984dq}
G.W.~Semenoff,
Phys. Rev. Lett. \textbf{53} (1984), 2449.

\bibitem{Hou:2006qc}
C.Y.~Hou, C.~Chamon and C.~Mudry,
Phys. Rev. Lett. \textbf{98} (2007), 186809.

\bibitem{Haldane:1988zza}
F.D.M.~Haldane,
Phys. Rev. Lett. \textbf{61} (1988), 2015-2018.

\bibitem{Gomes:1990ed}
M.~Gomes, R.S.~Mendes, R.F.~Ribeiro and A.J.~da Silva,
Phys. Rev. D \textbf{43} (1991), 3516-3523.

\bibitem{Hands:1994kb}
S.~Hands,
Phys. Rev. D \textbf{51} (1995), 5816-5826.

\bibitem{Hikami:1976at}
S.~Hikami and T.~Muta,
Prog. Theor. Phys. \textbf{57} (1977), 785-796.

\bibitem{Itoh:1994cr}
T.~Itoh, Y.~Kim, M.~Sugiura and K.~Yamawaki,
Prog. Theor. Phys. \textbf{93} (1995), 417-440.

\bibitem{Sugiura:1996xk}
M.~Sugiura,
Prog. Theor. Phys. \textbf{97} (1997), 311-326.

\bibitem{Kondo:1995np}
K.I.~Kondo,
Nucl. Phys. B \textbf{450} (1995), 251-266.

\bibitem{Hong:1993qk}
D.K.~Hong and S.H.~Park,
Phys. Rev. D \textbf{49} (1994), 5507-5511.

\bibitem{Ulybyshev:2013swa}
M.V.~Ulybyshev, P.V.~Buividovich, M.I.~Katsnelson and M.I.~Polikarpov,
Phys. Rev. Lett. \textbf{111} (2013), 056801.

\bibitem{Gies:2010st}
H.~Gies and L.~Janssen,
Phys. Rev. D \textbf{82} (2010), 085018.

\bibitem{DelDebbio:1997dv}
L.~Del Debbio, S.~Hands and J.C.~Mehegan,
Nucl. Phys. B \textbf{502} (1997), 269-308.

\bibitem{Burden:1986by}
C.~Burden and A.N.~Burkitt,
Europhys. Lett. \textbf{3} (1987), 545.

\bibitem{Kim:1996xza}
S.~Kim and Y.~Kim,
doi:10.1007/978-94-011-5812-1\_46
[arXiv:hep-lat/9605021 [hep-lat]].

\bibitem{Alexandru:2016ejd}
A.~Alexandru, G.~Basar, P.F.~Bedaque, G.W.~Ridgway and N.C.~Warrington,
Phys. Rev. D \textbf{95} (2017) no.1, 014502.

\bibitem{Narayanan:2021rcu}
R.~Narayanan,
[arXiv:2102.11367 [hep-lat]].

\bibitem{Christofi:2007ye}
S.~Christofi, S.~Hands and C.~Strouthos,
Phys. Rev. D \textbf{75} (2007), 101701.

\bibitem{Chandrasekharan:2011mn}
S.~Chandrasekharan and A.~Li,
Phys. Rev. Lett. \textbf{108} (2012), 140404.

\bibitem{Chandrasekharan:2013aya}
S.~Chandrasekharan and A.~Li,
Phys. Rev. D \textbf{88} (2013), 021701.

\bibitem{Hands:1992be}
S.~Hands, A.~Koci\'c and J.B.~Kogut,
Annals Phys. \textbf{224} (1993), 29-89.

\bibitem{Appelquist:1986fd}
T.W.~Appelquist, M.J.~Bowick, D.~Karabali and L.C.R.~Wijewardhana,
Phys. Rev. D \textbf{33} (1986), 3704.

\bibitem{Appelquist:1999hr}
T.~Appelquist, A.G.~Cohen and M.~Schmaltz,
Phys. Rev. D \textbf{60} (1999), 045003.

\bibitem{Drell:1976mj}
S.D.~Drell, M.~Weinstein and S.~Yankielowicz,
Phys. Rev. D \textbf{14} (1976), 1627.

\bibitem{Karsten:1979wh}
L.H.~Karsten and J.~Smit,
Phys. Lett. B \textbf{85} (1979), 100-102.

\bibitem{Wellegehausen:2017goy}
B.H.~Wellegehausen, D.~Schmidt and A.~Wipf,
Phys. Rev. D \textbf{96} (2017), 094504.

\bibitem{Kaplan:1992bt}
D.B.~Kaplan,
Phys. Lett. B \textbf{288} (1992), 342-347.

\bibitem{Ginsparg:1981bj}
P.H.~Ginsparg and K.G.~Wilson,
Phys. Rev. D \textbf{25} (1982), 2649.

\bibitem{Hands:2015qha}
S.~Hands,
JHEP \textbf{09} (2015), 047.

\bibitem{Neuberger:1997fp}
H.~Neuberger,
Phys. Lett. B \textbf{417} (1998), 141-144.

\bibitem{Neuberger:1998wv}
H.~Neuberger,
Phys. Lett. B \textbf{427} (1998), 353-355.

\bibitem{Hands:2015dyp}
S.~Hands,
Phys. Lett. B \textbf{754} (2016), 264-269.

\bibitem{Furman:1994ky}
V.~Furman and Y.~Shamir,
Nucl. Phys. B \textbf{439} (1995), 54-78.

\bibitem{Hands:2016foa}
S.~Hands,
JHEP \textbf{11} (2016), 015.

\bibitem{Vranas:1999nx}
P.~Vranas, I.~Tziligakis and J.B.~Kogut,
Phys. Rev. D \textbf{62} (2000), 054507.

\bibitem{Hands:2018vrd}
S.~Hands,
Phys. Rev. D \textbf{99} (2019), 034504.

\bibitem{Hands:2020itv}
S.~Hands, M.~Mesiti and J.~Worthy,
Phys. Rev. D \textbf{102} (2020), 094502.

\bibitem{Hernandez:1998et}
P.~Hern\'andez, K.~Jansen and M.~L\"uscher,
Nucl. Phys. B \textbf{552} (1999), 363-378.

\bibitem{Lenz:2019qwu}
J.J.~Lenz, B.H.~Wellegehausen and A.~Wipf,
Phys. Rev. D \textbf{100} (2019), 054501.

\bibitem{Gehring:2015vja}
F.~Gehring, H.~Gies and L.~Janssen,
Phys. Rev. D \textbf{92} (2015), 085046.

\bibitem{Schmidt:2016rtz}
D.~Schmidt, B.~Wellegehausen, A.~Wipf,
PoS \textbf{LATTICE2016} (2016), 247.

\bibitem{BenPage}
B.T.~Page, M.Phys. dissertation, Swansea University (2020).

\end{thebibliography}


\end{document}